\documentstyle[12pt,rotate]{article}

\input epsf.sty
\newcommand{\inclfig}[2]{\mbox{\epsfxsize=#1cm \epsfbox{#2.ps}}}
\newcommand{\insertfig}[2]{\mbox{\epsfxsize=#1cm \epsfbox{#2.eps}}}
\newcommand{\g}{{\sl g}}

\newcommand{\cJ}{{\cal J}}
\newcommand{\cZ}{{\cal Z}}

\newcommand{\cD}{{\cal D}}
\newcommand{\cK}{{\cal K}}

\newcommand{\cO}{{\cal O}}
\newcommand{\cH}{{\cal H}}
\newcommand{\cM}{{\cal M}}
\newcommand{\cP}{{\cal P}}
\newcommand{\cN}{{\cal N}}
\newcommand{\cQ}{{\cal Q}}
\newcommand{\cF}{{\cal F}}
\newcommand{\cE}{{\cal E}}
\font\cmss=cmss12 
\def\1{\hbox{{1}\kern-.25em\hbox{l}}}
\def\bfZ{\relax{\hbox{\cmss Z\kern-.4em Z}}}

\begin{document}
\begin{titlepage}

\centerline{\large \bf Integrability and WKB solution of twist-three
                       evolution equations.}

\vspace{10mm}

\centerline{\bf A.V. Belitsky\footnote{Alexander von Humboldt Fellow.}}

\vspace{10mm}

\centerline{\it Institut f\"ur Theoretische Physik, Universit\"at
                Regensburg}
\centerline{\it D-93040 Regensburg, Germany}

\vspace{30mm}

\centerline{\bf Abstract}

\hspace{0.8cm}

We identify an integrable one-dimensional inhomogeneous three-site open
spin chain which arises in the problem of diagonalization of twist-three
quark-gluon evolution equations in QCD in the chiral-odd sector. Making
use of the existence of a non-trivial `hidden' integral of motion the
problem of diagonalization of the evolution kernels is reduced to the
study of a second order finite-difference equation which is solved in WKB
approximation for large conformal spins of the three-particle system.
The energies (alias anomalous dimensions) of eigenstates with different
scale dependence are found in limiting cases and compared with numerical
calculations.

\vspace{6cm}


\vspace{0.5cm}


\end{titlepage}

\section{Brief phenomenological profile and outline.}

Deep inelastic scattering of leptons off nucleons, $\ell N \to \ell' X$,
served for a long time as the only and the most reliable piece of
knowledge about the partonic structure of hadrons in high energy
collisions via measurements of non- and polarized chiral-even structure
functions \cite{LamRey98}. However the feasibility of measurements of
other inclusive characteristics of hadrons was strongly limited. Nowadays
in order to study in detail the high-energy dynamics of constituents,
especially in polarized targets, one is forced to use complementary
sources of experimental information. In order to circumvent the above
disadvantages of conventional deep inelastic scattering process one has
to resort to nucleon-nucleon collisions, like the Drell-Yan lepton pair
production $h h' \to (\gamma^*, W, Z) \to \ell^+ \ell^- X$, which are not
affected by the above restraints and allow to explore a wealth of
hadronic properties, e.g.\ to access a new type of parton densities ---
chiral odd distributions $h_1$ at twist-2, $h_L$ and $e$ at twist-3, and,
related by QCD equation of motion to the latter two, generalized
three-parton correlators \cite{JafJi91}. The twist-3 distributions being
a direct manifestation of the quark-gluon interaction in non-perturbative
domain could serve as a testing ground for the phenomenological models
of confinement and, perhaps, future lattice simulations.

Similar to the more familiar chiral-even transverse spin structure
function $g_2$, $h_L$ although being of higher twist turns out to be
unique by its feature to appear as a leading unsuppressed effect in
longitudinal-transverse asymmetry in $p_{\|}p_\perp \to \ell^+ \ell^- X$
reaction \cite{JafJi91,KanKoiNis98}. However, an inevitable complication
of all experiments is that the data is taken at different values of a
hard momentum transfer $Q$ and, therefore, to deduce the shape of
distributions or to check sum rules one needs to know the way to
describe the variation of quantities in question when going up(down)wards
with $Q$ from the point of view of the underlying microscopic theory of
strong interaction --- Quantum Chromodynamics. Within the latter
framework this change is described by evolution equations which in
turn arise as a result of intrinsic ultraviolet divergencies of the
product of local elementary fields, --- which define an operator content
of distributions, --- separated by light cone distances, and thus
require renormalization which introduces a scale dependence. The specifics
of twist-3 sector as compared to the well-studied twist-2 case is that
the above mentioned structure functions being defined as two-particle
hadron-hadron matrix elements receive contributions from the three-parton
quark-gluon correlators \cite{KodTan98,Bel97}. Resolving this complication
in favour of a study of the renormalization group evolution of the
latter one ends up with a very complicated problem of working out the
mixing of three-particle local operators which in general cannot
be handled analytically.

One can ague that it is not really possible to see this logarithmic
violation of scaling experimentally due to the contamination of a more
strong $1/Q$ background behaviour of prefactors which accompany the
asymmetries. Nevertheless this knowledge is important for contrasting
the theoretical models with the measurable quantities. Since the model
estimations are usually done at a rather low momentum scale the
evolution effects are very prominent and change in a sizable way the
shape of an input distribution during short evolution times. So that
perturbative evolution can be thought of as a prime mover of gross
features of non-perturbative functions at hadronic scales, however
much larger then $\Lambda_{\rm QCD}$. Recall in this respect the
radiatively generated parton densities \cite{GRV}.

The distribution functions $h_L$ and $e$ as measured in most of
processes are given, up to `kinematical' Wandzura-Wilczek-type
contribution in $h_L$ which obeys the ordinary twist-2 evolution,
by particular integrals (see Eqs.\ (\ref{etoZ},\ref{hLtoZ}) below)
of three-particle quark-gluon correlation functions. The first
drastic simplification of their evolution occurs for large number
of colours. In multicolour limit the weight in the integrand of
convolution integrals coincides with the eigenfunction of the leading
order evolution kernel and, therefore, the above mentioned chiral-odd
structure functions evolve multiplicatively --- the so-called
`decoupling' phenomenon \cite{ABH91,BBKT97,Bel97a}. However, in the
spin-independent Drell-Yan process the three-particle correlators
associated with the unpolarized distribution $e$ enter folded with
different coefficient functions \cite{JafJi91} and, therefore, will
not evolve autonomously even in multicolour limit. In the realistic
QCD case, when the $1/N_c^2$ correction are accounted for, the
structure of evolution becomes more complicated: all three-particle
operators are of relevance. But the violation of the simple pattern
of multicolour evolution can be well understood once it is the case for
large-$N_c$ limit. Since these corrections produce $\propto (10-20) \%$
effects in the evolution of the model distributions\footnote{We discard
from consideration some pathological shapes of gluon distribution in
three-parton correlators which induce non-smooth evolution
\cite{BelMul97}.} which although being small and apparently might
be sufficient for practical purposes at present should still be
under control. This paper is devoted to an analysis of the
large-$N_c$ spectrum of the three-particle evolution equations.

Our consequent presentation is organized as follows: In the next
section we describe an approach to study the scale violation of
the so-called quasi-partonic operators which are related to the
quantities in question by QCD equations of motion. We show that
the tree-level conformal symmetry of classical QCD Lagrangian
allows for a partial diagonalization of the evolution kernels. In
section 3 we elaborate on this point further and construct an
appropriate three-particle conformal basis. A distinguishing
property of the problem at hand is its equivalence to an exactly
solvable one-dimensional open spin chain which is established in
section 4. Making use of the existence of a `hidden' integral of
motion allows to reduce the eigenvalue problem for the energy of
the system (alias anomalous dimensions) to the solution of a
second order finite-difference equation --- referred to as the
master equation --- which is derived in section 5. Consequent
discussion concerns the study of the analytical structure of the
spectrum of the conserved charge for large values of the total
conformal spin. Its asymptotic solutions are given in section 6.
These results allow to describe fairly well the spectrum of energy
eigenvalues found by numerical diagonalization of the mixing matrix
of the anomalous dimensions. Finally, we go to conclusions.

\section{Approach to scaling violation phenomena.}

In QCD the twist-3 distributions, $h_L$ and $e$, are expressed via
the light-cone Fourier transformation of bilocal quark string operators
\cite{JafJi91}
\begin{equation}
\label{DefinitEH}
\left( { e(x) \atop h_L (x) } \right)
=\frac{x}{2}\int \frac{d \kappa}{2\pi}
e^{i \kappa x}
\langle h | \bar \psi (0) \Phi [0, \kappa n]
\left(  {1 \atop i \sigma_{+-} }\right)
\psi (\kappa n)|h \rangle ,
\end{equation}
with a phase factor $\Phi [0, \kappa n]$ which ensures the non-abelian
gauge invariance. Although these definitions resemble the ones of ordinary
twist-2 parton densities a more closer inspection immediately reveals
an essential difference from the latter: twist-3 functions involve
the ``bad'' light-cone components of the fields in the language of
the Kogut-Soper infinite momentum frame formalism \cite{KogSop70}.
By virtue of the QCD Heisenberg equation of motion for the quark field
and constraints coming from Lorentz invariance, the functions
(\ref{DefinitEH}) can be rewritten in an explicitly interaction
dependent way as\footnote{The first contribution on the r.h.s.\ of
Eq.\ (\ref{hLtoZ}) is afore mentioned `kinematical' twist-2 WW-term
\cite{JafJi91}.}
\cite{BelMul97}
\begin{eqnarray}
\label{etoZ}
e (x)
&=& \int \frac{d \beta}{x - \beta} Z (x, \beta) ,\\
\label{hLtoZ}
h_L (x)
&=& 2 x^2 \int_{x}^{1} \frac{d \beta}{\beta^2} h_1 ( \beta )
+ x^2 \int_{x}^{1} \frac{d \beta}{\beta^2}
\int \frac{d \beta'}{ \beta' - \beta }
\left[ \frac{\partial}{\partial \beta}
- \frac{\partial}{\partial \beta'} \right]
\widetilde Z ( \beta, \beta' ).
\end{eqnarray}
in terms of correlation functions
\begin{equation}
\label{CorrelFunct}
Z ( x_1, x_2 = - x_1 - x_3, x_3 ) =
\int \frac{d \kappa_1}{2 \pi} \frac{d \kappa_3}{2 \pi}
e^{i \kappa_1 x_1 + i \kappa_3 x_3}
\langle h |
\frac{1}{2}
\left\{
\cZ (\kappa_1, 0, \kappa_3) \pm
\cZ ( - \kappa_3, 0, - \kappa_1)
\right\}
| h \rangle
\end{equation}
of the non-local quark-gluon operators
\begin{equation}
\label{oddOperator}
\left(
\begin{array}{c}
\cZ \\
\widetilde\cZ
\end{array}
\right)
(\kappa_1, \kappa_2, \kappa_3)
= \frac{1}{2} \bar\psi (\kappa_3 n) \sigma^\perp_{\mu +}
\left(
\begin{array}{c}
1 \\
\gamma_5
\end{array}
\right)
t^a \g G^{a \perp}_{+ \mu} (\kappa_2 n) \psi (\kappa_1 n) ,
\end{equation}
with gauge link factors omitted for brevity. However the operators
in Eq.\ (\ref{oddOperator}), while contain one parton more, share a
bulk of properties with the usual twist-2 operators: only ``good''
light-cone components and collinear momenta appear. ``Bad'' fields
and transverse momentum dependence have been eliminated in favour of
interaction which originates them. Thus in this description the theory
is endowed again with a parton-like interpretation albeit of the QCD
improved quark model. And these are the properties which distinguish
a special class of the so-called quasi-partonic operators \cite{BFKL85}
which are the most appropriate for the study of renormalization group
dependence as the corresponding evolution equation turns out to be
homogeneous. This latter property has arisen due to the fact that the
pairs of parton fields in Eq.\ (\ref{oddOperator}) correspond to the
familiar twist-2 operators although they can have fermionic quantum
numbers besides bosonic.

\subsection{Evolution equation.}

\begin{figure}[t]
\begin{center}
\vspace{-0.3cm}
\hspace{1cm}
\mbox{
\begin{picture}(0,220)(270,0)
\put(0,-30){\inclfig{18}{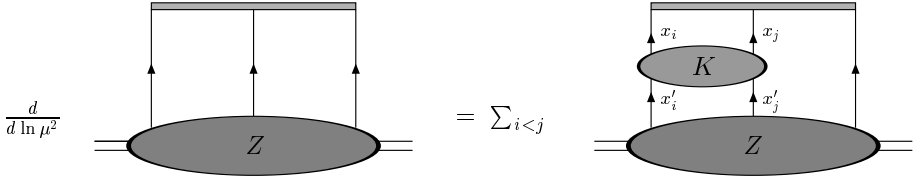}}
\end{picture}
}
\end{center}
\vspace{-6cm}
\caption{\label{DiagramEvEq} The diagrammatical representation of the
generic form of the leading order evolution equation for the three-parton
correlator $Z (x_1, x_2, x_3)$ in the light-cone gauge. The blob $K$
stands for the pair-wise kernel of interaction of $t$-channel particles.}
\end{figure}

The generic form of the evolution equation for twist-3 correlation
functions in the momentum fraction formalism looks like (see
Fig.\ \ref{DiagramEvEq})
\begin{equation}
\label{EvolEq}
\mu^2 \frac{d}{d \mu^2} Z (x_1, x_2, x_3)
\!=\! \int \prod_{i=1}^{3} d x'_i \,
\delta \left( \sum_{i = 1}^{3} x'_i \right)
\mbox{\boldmath$K$} \left( \{ x_i \} | \{ x'_i \} \right)
Z (x'_1, x'_2, x'_3) ,
\end{equation}
with the constraint $x_1 + x_2 + x_3 = 0$ of the forward inclusive
kinematics for the momentum fractions of partons imposed on the both
sides of this equation. Due to topology of one-loop Feynman diagrams
the evolution kernel $\mbox{\boldmath$K$}$ has a simple pair-wise
structure at leading order:
\begin{equation}
\label{LOkernel}
\mbox{\boldmath$K$} \left( \{x_i\} | \{x'_i\} \right) = \sum_{i < j}
K \left( x_i, x_j | x'_i, x'_j \right)
\delta \left( x_i + x_j - x'_i - x'_j \right) ,
\end{equation}
with $K \left( x_i, x_j | x'_i, x'_j \right)$ being a twist-2
interaction kernel of two nearby particles.

The solution to the leading order evolution equation (\ref{EvolEq})
can be written in the form
\begin{equation}
Z ( x_1, x_2, x_3 | Q^2 )
= \sum_{\{ \alpha \}}
{\mit\Psi}_{\{ \alpha \}} (x_1, x_2, x_3)
\left(
\frac{\alpha_s ( Q_0^2 )}{\alpha_s ( Q^2 )}
\right)^{2 \cE_{\{ \alpha \}} / \beta_0}
\langle\langle \cZ_{\{ \alpha \}} ( Q_0^2 ) \rangle\rangle
\end{equation}
of expansion w.r.t.\ polynomials ${\mit\Psi}_{\{ \alpha \}}
(x_1, x_2, x_3)$ which are the eigenfunctions of the leading order
kernel (\ref{LOkernel}) with eigenvalues $\cE_{\{ \alpha \}}$ and
where $\langle\langle \cZ_{\{ \alpha \}} \rangle\rangle$ denote
reduced hadronic matrix elements of local operators defined as
(Mellin-like) moments of the three-parton correlation function.
The summation goes over a set of quantum numbers $\{ \alpha \}$
which enumerate the solutions, e.g.\ spin of the system etc. And
$\beta_0 = \frac{4}{3} T_F N_f - \frac{11}{3} C_A$ is the first
coefficients of the QCD Gell-Mann-Low function. Since the polynomials
depending on the momentum fractions can be immediately translated by
Fourier transform to the ones depending on derivatives acting
on the elementary field operators, they will correspond to the
eigenvectors of the anomalous dimension matrix in a basis of local
operators. Thus the problem of solving the evolution equation is
equivalent to the diagonalization of the matrix of anomalous
dimensions
\begin{equation}
\label{EigenProblem}
\mbox{\boldmath$K$} {\mit\Psi}_{\{ \alpha \}}
= - \frac{\alpha_s}{2 \pi} \cE_{\{ \alpha \}}
{\mit\Psi}_{\{ \alpha \}} ,
\end{equation}
where we have explicitly extracted the dependence on the coupling
constant $\alpha_s$.

For the problem at hand the total evolution kernel takes the
same form for the parity-even ($e$) and -odd ($h_L$) sectors
\begin{eqnarray}
\label{TotalKernelOdd}
\mbox{\boldmath$K$} \left( \{x_i\} | \{x'_i\} \right)
= - \frac{\alpha_s}{2 \pi}
\biggl\{
&&\!\!\!\!\!\!\!\!\!\!
{^{q \bar q}\! K^T_{(8)}} \left( x_1, x_3 | x'_1, x'_3 \right)
\delta (x_1 + x_3 - x'_1 - x'_3) \\
+
&&\!\!\!\!\!\!\!\!\!\!
{^{q g}\! K^V_{(3)}} \left( x_1, x_2 | x'_1, x'_2 \right)
\delta (x_1 + x_2 - x'_1 - x'_2) \nonumber\\
+
&&\!\!\!\!\!\!\!\!\!\!
{^{q g}\! K^V_{(3)}} \left( x_3, x_2 | x'_3, x'_2 \right)
\delta (x_3 + x_2 - x'_3 - x'_2)
- \frac{\beta_0}{4} \delta (x_1 - x'_1) \delta (x_2 - x'_2)
\biggr\} , \nonumber
\end{eqnarray}
with known non-forward evolution kernels, ${^{\phi_1 \phi_2}\! K^D_{(c)}}$,
for colour triplet ($c = 3$) vector ($D = V$) and colour octet ($c = 8$)
transversity ($D = T$) sectors. The selection of particular solutions
comes from different parity properties of three-variable functions $Z$
and $\widetilde Z$ w.r.t.\ permutation of quark fields: the former is
symmetrical $Z(x_1, x_2) = Z (x_2, x_1)$ while the latter is
antisymmetrical $\widetilde Z(x_1, x_2) = - \widetilde Z (x_2, x_1)$
under this interchange. We have added in Eq.\ (\ref{TotalKernelOdd}) the
trivial charge renormalization counterterm which is traced back to the
presence of the coupling constant in the definition of the composite
operator (\ref{oddOperator}).

\subsection{Eigenvalue problem.}

Due to absence of the QCD $\beta$-function contribution in leading order
pair-wise evolution kernels we can use the conformal invariance
property of QCD at tree level since the renormalization counterterms
of composite operators at leading order are just the ones in classical
theory while from the side of QCD Lagrangian the only counterterms
which are required in the physical sector to cure the divergencies
are the field renormalization constants. But the anomalous dimensions
associated to them can be embedded into redefined representations of
the conformal group --- replacing the canonical by the scale field
dimensions --- and thus preserve conformal covariance.

Making use of this property of classical QCD one can diagonalize the
pair-wise non-forward kernels \cite{EfrRad79}-\cite{BelMul99a} and
present them in a factorized form \cite{BelMul98a}
\begin{equation}
\label{FactorizedKernel}
{^{\phi_1 \phi_2}\! K} (x_1, x_2 | x'_1, x'_2)
= \frac{1}{2}
\sum_{j = 0}^{\infty}
\frac{w (x_1, \nu_1 | x_2, \nu_2)}{\omega_j (\nu_1, \nu_2)}
P^{\left( \nu_2, \nu_1 \right)}_j
\left( \frac{x_1 - x_2}{x_1 + x_2} \right)
{^{\phi_1 \phi_2}\! \gamma_j}\,
P^{\left( \nu_2, \nu_1 \right)}_j
\left( \frac{x'_1 - x'_2}{x'_1 + x'_2} \right),
\end{equation}
where $w (x_1, \nu_1 | x_2, \nu_2) = x_1^{\nu_1} x_2^{\nu_2}$
and $\omega_j (\nu_1, \nu_2) = \frac{\Gamma (j + \nu_1 + 1)
\Gamma (j + \nu_2 + 1)}{\left( 2j + \nu_1 + \nu_2 + 1 \right) j!
\Gamma (j + \nu_1 + \nu_2 + 1)}$, with $\nu_\ell = d_\ell + s_\ell
- 1$, and $d_\ell$ and $s_\ell$ standing for the canonical
dimension and spin of the constituent $\phi_\ell = \{q, g\}$.
The $P^{\left( \nu_2, \nu_1 \right)}_j$ are the usual Jacobi
polynomials \cite{BatErd53v2} which form an infinite dimensional
representation of the collinear conformal group \cite{MacSal69},
--- whose $su(1,1)$ algebra
\begin{equation}
[\cJ^3 , \cJ^\pm]_- = \pm \cJ^\pm ,
\qquad
[\cJ^+ , \cJ^-]_- = -2 \cJ^3 ,
\end{equation}
is defined by generators of the momentum $\cJ^+ = i \cP_+$,
special conformal transformation $\cJ^- = \frac{i}{2} \cK_-$ and
particular combination of dilatation and angular momentum
$\cJ^3 = \frac{i}{2} (\cD + \cM_{-+})$, --- in the space spanned
by bilinear (in elementary field operators) local operators with
total derivatives \cite{EfrRad79}-\cite{Ohr82}
\begin{equation}
\label{ConformOper}
\cO_{jk} = \phi_2 (i \partial_1 + i \partial_2)^k
P^{(\nu_2, \nu_1)}_j
\left( \frac{\partial_1 - \partial_2}{\partial_1 + \partial_2} \right)
\phi_1 .
\end{equation}
These operators possess conformal spin $J_{12} = j +
\frac{1}{2}(\nu_1 + \nu_2 + 2 )$ which is the eigenvalue
$[\mbox{\boldmath$\cJ$}^2, \cO_{jk}]_- = J_{12}(J_{12} - 1)
\cO_{jk}$ of the Casimir operator $\mbox{\boldmath$\cJ$}^2
= \cJ^3 ( \cJ^3 - 1 ) - \cJ^+ \cJ^-$.

In the Eq.\ (\ref{FactorizedKernel}), all dynamics of the system
is concentrated in the anomalous dimensions of the conformal
operators, ${^{\phi_1 \phi_2}\! \gamma_j}$. Thus the former depend
only on the eigenvalues of the two-particle Casimir operators in
subchannels. For the eigenvalue problem (\ref{EigenProblem}) it can
be translated to the fact that the pair-wise kernels commute with
generators of conformal transformation, thus, again being a function
of Casimir operator only.

This dependence can be easily traced from the corresponding eigenvalues.
For our consequent considerations we need the bosonic colour-octet
chiral-odd, ${^{q \bar q} \gamma^T_{(8)}}_j$ and fermionic
colour-triplet chiral-even\footnote{We do not bother about parity since
it is irrelevant for fermionic operators.} ${^{q g} \gamma^V_{(3)}}_j$
anomalous dimensions which read
\begin{eqnarray}
\label{AnomalousDimensions}
{^{q \bar q} \gamma^T_{(8)}}_j
&=& \left( C_F - \frac{C_A}{2} \right)
\left( 4 \psi (j + 2) - 4 \psi (1) - 3 \right), \\
{^{q g} \gamma^V_{(3)}}_j
&=&
\frac{C_A}{2}
\left(
2 \psi (j + 1) + 2 \psi (j + 4) - 4 \psi (1) \right) \nonumber\\
&&\hspace{4cm} + \left( C_F - \frac{C_A}{2} \right)
\frac{4 \sigma (j)}{(j + 1 )(j + 2)(j + 3)}
+ \frac{1}{4} \left( \beta_0 - 3 C_A \right) ,
\end{eqnarray}
where $\sigma (j) = (- 1)^j$.

Therefore, the evolution equation can be reformulated into the
eigenvalue problem for the three-particle system in a basis of
local operators
\begin{equation}
\label{Eigensystem}
\cH_{\rm QCD}\, {\mit\Psi} = \cE_{\rm QCD}\, {\mit\Psi} ,
\end{equation}
where $\mbox{\boldmath$K$} \equiv - \frac{\alpha_s}{2 \pi} \cH$ with
Hamiltonian
\begin{equation}
\label{FiniteNcHamiltonian}
\cH_{\rm QCD}
= {^{q \bar q}\!\cH^T_{(8)}} (\hat J_{13})
+ {^{q g}\!\cH^V_{(3)}} (\hat J_{12})
+ {^{q g}\!\cH^V_{(3)}} (\hat J_{32})
- \frac{\beta_0}{4} ,
\end{equation}
and we have defined ${^{q \bar q}\!\gamma}_{J - 2} =
2\, {^{q \bar q}\!\cH} (J)$ and ${^{q g}\!\gamma}_{J - 5/2}
= 2\, {^{q g}\!\cH} (J)$. The operators $\hat J$'s are formally
determined as solutions of the operator equation\footnote{Note that
the operators $\hat J$ introduced here differs from the ones defined
in Ref.\ \cite{Bel99a} by a shift: $\hat J^{\rm here} =
\hat J^{\cite{Bel99a}} + 1$.} for Casimir operator in some
representation\footnote{Once conformal invariance is established
we can abstract from the particular representation used for the
derivation.} ${\mbox{\boldmath$\hat J$}}^2 = \hat J (\hat J - 1)$.

One of the complications to find the solution of the eigenvalue problem
(\ref{Eigensystem}) is a rather non-trivial dependence on the colour
structure of the quark-gluon system. In order to get rid of it we take
large $N_c$ limit in Eq.\ (\ref{FiniteNcHamiltonian}) which results
into omission of the $1/N_c^2$ terms in Eq.\ (\ref{AnomalousDimensions}).
This leads to the Hamiltonian
\begin{equation}
\label{LargeNcHamiltonian}
\cH
= \frac{N_c}{2}
\left\{ h_{1 2} + h_{2 3} - \frac{3}{2} \right\} ,
\qquad\mbox{with}\qquad
h_{a b} = \psi \left( \hat J_{a b} - \frac{3}{2} \right)
+ \psi \left( \hat J_{a b} + \frac{3}{2} \right)
- 2 \psi (1) ,
\end{equation}
whose diagonalization will be our cherished goal.

\section{Conformal basis: General formalism.}

As we have seen above the conformal covariance provides a partial
diagonalization of the three-parton evolution kernels. Therefore,
an appropriate basis constructed with regards to the tree-level
conformal symmetry can simplify the problem further. This question
is considered below in the present section.

\subsection{Auxiliary $\theta$-space.}

It turns out to be convenient to introduce an auxiliary vector
space where the conformal invariance properties are reformulated
in a manifest and simple way. This description appears once we
consider the elements
\begin{equation}
\label{ThetaToDeriv}
\theta^k \equiv \frac{\partial^k_+ \phi}{\Gamma (k + \nu + 1)} ,
\end{equation}
as building blocks of the formalism. In this representation
$[\cJ^{\pm,3} ,\chi ( \theta )]_- = \hat J^{\pm,3} \chi ( \theta )$
the generators take the following form
\begin{equation}
\hat J^+ = (\nu + 1) \theta + \theta^2 \frac{\partial}{\partial\theta},
\quad
\hat J^- = \frac{\partial}{\partial\theta},
\quad
\hat J^3 = \frac{1}{2}(\nu + 1) + \theta \frac{\partial}{\partial\theta} ,
\end{equation}
and the quadratic Casimir operator is given by
\begin{equation}
{\mbox{\boldmath$\hat J$}}^2
= \hat J^3 ( \hat J^3 - 1 ) - \hat J^+ \hat J^- .
\end{equation}
For a multi-variable function $\chi ( \theta_1, \theta_2,\dots,\theta_n )$
the operators are defined by the sum of single particle ones as
$\hat J^{\pm,3} = \sum_{\ell = 1}^{n} \hat J^{\pm,3}_\ell$ and
they obey the usual commutation relations: $[\hat J^3 , \hat J^\pm]_-
= \pm \hat J^\pm$, $[\hat J^+ , \hat J^-]_- = -2 \hat J^3$. Obviously,
the single particle state $\theta_\ell$ is an eigenstate of the Casimir
operator
\begin{equation}
{\mbox{\boldmath$\hat J$}}^2_\ell \theta_\ell
=
{\mbox{\boldmath$\hat \jmath$}}^2_\ell \theta_\ell,
\qquad\mbox{with}\qquad
{\mbox{\boldmath$\hat \jmath$}}^2_\ell
\equiv \frac{1}{4}( \nu_\ell^2 - 1 ) .
\end{equation}

Since the Eq.\ (\ref{Eigensystem}) defines the physical anomalous
dimensions, the latter have to be real that requires the Hamiltonian
(\ref{LargeNcHamiltonian}) to be selfadjoint\footnote{To be precise
Hermitean matrices do possess real eigenvalues, however, the inverse
statement in not true. Recall in this respect triangular matrices with
real diagonal elements.} w.r.t. an appropriate scalar product. We
define the latter as
\begin{equation}
\label{ScalarProduct}
\langle \chi ( \theta_1, \theta_2,\dots,\theta_n ) |
\chi ( \theta_1, \theta_2,\dots,\theta_n ) \rangle
= \int_{{\mit\Omega}} d \cM \
\chi ( \bar\theta_1, \bar\theta_2,\dots,\bar\theta_n )
\chi ( \theta_1, \theta_2,\dots,\theta_n ) ,
\end{equation}
where
\begin{eqnarray*}
d \cM \equiv
\prod_{\ell = 1}^{n}
\frac{d \theta_\ell d \bar\theta_\ell}{2 \pi i}
( 1 - \theta_\ell \bar\theta_\ell )^{\nu_\ell - 1}
\qquad\mbox{and}\qquad
{\mit\Omega} = \bigcup^n_{\ell = 1} \{ |\theta_\ell| \leq 1 \}
\end{eqnarray*}
and $\bar\theta = \theta^\ast$. Then it can be seen that the Casimir
operator and, as a consequence, the Hamiltonian are selfadjoint
operators w.r.t.\ such defined inner product
\begin{eqnarray*}
( {\mbox{\boldmath$\hat J$}}^2 )^\dagger
= {\mbox{\boldmath$\hat J$}}^2,
\qquad
\cH^\dagger = \cH ,
\end{eqnarray*}
and therefore ${\rm Im}\, \cE = 0$.

\subsection{Fundamental basis.}

Using the above definitions one can construct a second order
differential operator (here and throughout $\theta_{ab} \equiv
\theta_a - \theta_b$)
\begin{equation}
{\mbox{\boldmath$\hat J$}}^2_{ab}
= - \theta_{ab}^{1 - ( \nu_a + \nu_b )/2}
\partial_a \partial_b
\theta_{ab}^{1 + ( \nu_a + \nu_b )/2}
+ \frac{1}{2} ( \nu_b - \nu_a ) \theta_{ab}
\left( \partial_a + \partial_b \right) ,
\end{equation}
which corresponds to the two-particle Casimir operator in
$(a,b)$-subchannel. Its eigenstates $\left( {\hat J}^+_{12} \right)^k
\theta_{12}^j$ coincide with bilinear conformal operators
(\ref{ConformOper}) and possess the same eigenvalues. The
three-particle Casimir is an obvious generalization
\begin{equation}
\label{ThreeCasimir}
{\mbox{\boldmath$\hat J$}}^2 =
{\mbox{\boldmath$\hat J$}}^2_{12}
+ {\mbox{\boldmath$\hat J$}}^2_{23}
+ {\mbox{\boldmath$\hat J$}}^2_{13}
- \sum_{\ell = 1}^{3} {\mbox{\boldmath$\hat \jmath$}}^2_\ell .
\end{equation}

As it has been advocated in Refs.\ \cite{BraDerMan98}-\cite{Bel99a} it
proves convenient to introduce the following basis of three-particle
operators which simultaneously diagonalize the total three-particle
Casimir operator ${\mbox{\boldmath$\hat J$}}^2$ and one in a
subchannel ${\mbox{\boldmath$\hat J$}}_{ab}^2$, say $a = 1$ and
$b = 2$,
\begin{eqnarray}
\label{DiagTot}
{\mbox{\boldmath$\hat J$}}^2
\cP_{J; j} ( \theta_1, \theta_2 | \theta_3 )
\!\!\!&=&\!\!\!
\left( J + \frac{1}{2}(\nu_1 + \nu_2 + \nu_3 + 3 )\right)
\left( J + \frac{1}{2}(\nu_1 + \nu_2 + \nu_3 + 1 )\right)
\cP_{J; j} ( \theta_1, \theta_2 | \theta_3 ) , \\
\label{Diag12}
{\mbox{\boldmath$\hat J$}}^2_{12}
\cP_{J; j} ( \theta_1, \theta_2 | \theta_3 )
\!\!\!&=&\!\!\!
\left( j + \frac{1}{2}( \nu_1 + \nu_2) \right)
\left( j + \frac{1}{2}( \nu_1 + \nu_2 + 2 ) \right)
\cP_{J; j} ( \theta_1, \theta_2 | \theta_3 ) .
\end{eqnarray}
The states with unit norm w.r.t.\ the scalar product
(\ref{ScalarProduct}) are defined immediately from the
above equations and read
\begin{equation}
\label{ThreePointBasis}
\cP_{J; j} ( \theta_1, \theta_2 | \theta_3 )
=
\cN^{-1} (J, j | \nu_1, \nu_2, \nu_3)
\sum_{k = 0}^{J - j}
\frac{(-1)^k { J - j \choose k }}{( 2 j + \nu_1 + \nu_2 + 2 )_k}
\theta_3^{J - j - k}
\left( {\hat J}^+_{12} \right)^k \theta_{12}^j
\end{equation}
with normalization factor
\begin{equation}
\cN^2 (J, j | \nu_1, \nu_2, \nu_3)
=
\frac{ j! (J - j)! ( j + \nu_1 + \nu_2 + 1 )_j}{
( \nu_1 )_{j + 1} ( \nu_2 )_{j + 1} ( \nu_3 )_{J - j + 1} }
\frac{ ( J + j + \nu_1 + \nu_2 + \nu_3 + 2 )_{J - j} }{
( 2 j + \nu_1 + \nu_2 + 2 )_{J - j} } ,
\end{equation}
where we have introduced the Pochhammer symbol $( \alpha )_\ell
= \Gamma ( \alpha + \ell ) / \Gamma ( \alpha )$. By construction
the above basis is orthonormal
$\langle {\cal P}_{J';j'} ( \theta_1, \theta_2 | \theta_3 )|
{\cal P}_{J;j} ( \theta_1, \theta_2 | \theta_3 ) \rangle =
\delta_{J'J} \delta_{j'j}$.

\subsection{Matrix elements.}

The form of the basis given by Eq.\ (\ref{ThreePointBasis}) is
especially suitable for evaluation of matrix elements of $SU(1,1)$
generators. The `proper' generators of the basis $\cP (\theta_1,
\theta_2 | \theta_3)$ are diagonal (\ref{Diag12},\ref{DiagTot}) and
the only operators which possess off-diagonal elements are
${\mbox{\boldmath$\hat J$}}^2_{(1,2)3}$. The important property
of the basis (\ref{ThreePointBasis}) which distinguishes it among
other possibilities is that the matrix elements of `improper'
operators are three-diagonal only
\begin{equation}
\langle \cP_{J'; j'}
( \theta_1, \theta_2 | \theta_3 ) |
{\mbox{\boldmath$\hat J$}}^2_{23}
| \cP_{J; j} ( \theta_1, \theta_2 | \theta_3 ) \rangle
= \delta_{J'J}
\left(
\delta_{j'j} [ {\mbox{\boldmath$\hat J$}}^2_{23} ]_{j,j}
+
\delta_{j',j + 1} [ {\mbox{\boldmath$\hat J$}}^2_{23} ]_{j + 1,j}
+
\delta_{j',j - 1} [ {\mbox{\boldmath$\hat J$}}^2_{23} ]_{j - 1,j}
\right)
\end{equation}
where the diagonal part is
\begin{equation}
[ {\mbox{\boldmath$\hat J$}}^2_{23} ]_{j,j}
= \frac{1}{2}
\left(
[ {\mbox{\boldmath$\hat J$}}^2 ]_{jj}
-
[ {\mbox{\boldmath$\hat J$}}^2_{12} ]_{jj}
+
\sum_{\ell = 1}^{3}
{\mbox{\boldmath$\hat \jmath$}}^2_\ell
+
( {\mbox{\boldmath$\hat \jmath$}}^2_2
- {\mbox{\boldmath$\hat \jmath$}}^2_1 )
\left(
[ {\mbox{\boldmath$\hat J$}}^2 ]_{jj}
-
{\mbox{\boldmath$\hat \jmath$}}^2_3
\right)
[ {\mbox{\boldmath$\hat J$}}^2_{12} ]_{jj}^{- 1}
\right) ,
\end{equation}
and the non-diagonal elements are
\begin{equation}
[ {\mbox{\boldmath$\hat J$}}^2_{23} ]_{j + 1,j}
=
\frac{\cN (J, j | \nu_1, \nu_2, \nu_3)}{
\cN (J, j + 1 | \nu_1, \nu_2, \nu_3)}
(j + 1)(J - j + \nu_3) .
\end{equation}
From hermiticity it follows that
$[ {\mbox{\boldmath$\hat J$}}^2_{23} ]_{j + 1,j} =
[ {\mbox{\boldmath$\hat J$}}^2_{23} ]_{j,j + 1}$.

\subsection{Other representations of the basis.}

Since the operator (\ref{ThreePointBasis}) is the highest weight of
the conformal group in the three-particle basis, i.e.\ $\hat J^- \cP
= 0$, and is the homogeneous polynomial (through Eq.\ (\ref{DiagTot})) of
degree $J$ by the Euler theorem, this suggests that the net dependence
of Eq.\ (\ref{ThreePointBasis}) is on a single variable only (up to a
prefactor). This is really the case and immediately one can perform
the resummation of the series in Eq.\ (\ref{ThreePointBasis}) to
obtain a compact expression for the $\theta$-space basis (cf.\
\cite{BraDerKorMan99})
\begin{equation}
\label{HyperBasis}
\cP_{J; j} ( \theta_1, \theta_2 | \theta_3 )
= \cN^{-1} (J, j | \nu_1, \nu_2, \nu_3)
\theta_{12}^J \, \theta^{j - J}
{_2F_1} \left( \left.
{
j - J , j + \nu_1 + 1
\atop
2j + \nu_1 + \nu_2 + 2
}
\right| \theta \right) ,
\quad\mbox{with}\quad
\theta \equiv \frac{\theta_{12}}{\theta_{32}} .
\end{equation}
In general the solution to the condition $\hat J^- \cP = 0$ is
given by a translation invariant polynomial, e.g.\ $\cP \propto
\sum_{j_1 + j_2 = J} c_{j_1, j_2} \theta_{12}^{j_1} \theta_{32}^{j_2}$
with undetermined expansion coefficients. This is the standard
choice which leads to Appel polynomials when transformed back
to the space of local operators \cite{Ohr82}. If, as we have done
above, we will impose an additional constraint (\ref{Diag12}) we
get a recursion relation for $c$'s which when solved gives
(\ref{HyperBasis}).

The relation to the space of local operators can be readily figure
out by using the definition (\ref{ThetaToDeriv}) so that after some
algebra we end up with (cf.\ \cite{BraDerKorMan99})
\begin{eqnarray}
\label{ThreeLocal}
&&\phi_2
\left( i \partial_1 + i \partial_2
\right)^j
P_j^{( \nu_2, \nu_1 )}
\left(
\frac{i \partial_1 - i \partial_2}{i \partial_1 + i \partial_2}
\right)
\phi_1
\left( i \partial_1 + i \partial_2 + i \partial_3 \right)^{J - j}
P_{J - j}^{( 2j + \nu_1 + \nu_2 + 1, \nu_3 )}
\left(
\frac{i \partial_3 - i \partial_1 - i \partial_2}{i \partial_1
+ i \partial_2 + i \partial_3}
\right)
\phi_3 \nonumber\\
&&=
i^J \frac{\Gamma (j + \nu_1 + 1) \Gamma (j + \nu_2 + 1)}{\Gamma (j + 1)} \\
&&\qquad\qquad\qquad\qquad
\times
\frac{\Gamma (J - j + \nu_3 + 1)}{\Gamma (J - j + 1)}
\frac{\Gamma (J + j + \nu_1 + \nu_2 + 2)}{\Gamma
(2 j + \nu_1 + \nu_2 + 2)}
\cN (J, j | \nu_1, \nu_2, \nu_3)
\cP_{J; j} ( \theta_1, \theta_2 | \theta_3 ). \nonumber
\end{eqnarray}

\subsection{Racah decomposition.}

Since the choice of the second condition (\ref{Diag12}) on the basis
is arbitrary, --- one can analogously take a requirement for the
element of the basis to be an eigenstate of operator
${\mbox{\boldmath$\hat J$}}^2_{(1,2)3}$, --- one poses the question
as to how the transformation from one basis to another is
accomplished. Obviously, we meet a standard quantum mechanical
problem of addition of three `angular momenta'. According to
general theory they are related via the decomposition
\begin{equation}
P_{13} \cP_{J; j} ( \theta_1, \theta_2 | \theta_3 )
=
\cP_{J; j} ( \theta_3, \theta_2 | \theta_1 )
= \sum_{k = 0}^{J}
W_{jk} (J)
\cP_{J; k} ( \theta_1, \theta_2 | \theta_3 ) ,
\end{equation}
where the coefficients $W_{jk} (J)$ are viewed as Racah symbols
for $SU(1,1)$. The operator $P_{13}$ permutes the fields `1' and `3'
in teh operators $\cP_{J;j}$. The explicit form of $W_{jk} (J)$ can
be easily deduced from the above formula since\footnote{The $W_{jk} (J)$
are the matrix elements of the operator $P_{13}$.} $W_{jk} (J)
= \langle \cP_{J; k} ( \theta_1, \theta_2 | \theta_3 ) |
\cP_{J; j} ( \theta_3, \theta_2 | \theta_1 ) \rangle$ and making
use of the representation (\ref{ThreeLocal}) with the derivatives
loosely interpreted as momentum fractions. Going to the special
system with $i \partial_1 = z$, $i \partial_2 = 1 - z$ and
$i \partial_3 = 0$, where the calculation methods of Ref.\
\cite{BelMul99a} can be applied in a straightforward way, we end up
with the result
\begin{eqnarray}
W_{jk} (J)
&=&
(- 1)^{J - k}
\frac{\Gamma (J - j + 1)}{\Gamma (J - k + 1) \Gamma(k - j + 1)}
\frac{\Gamma (J + k + \nu_1 + \nu_2 + \nu_3 + 2)}{\Gamma
(J + j + \nu_1 + \nu_2 + \nu_3 + 2)} \nonumber\\
&\times&
\frac{\Gamma (k + \nu_1 + 1) \Gamma (k + \nu_2 + 1)
\Gamma (k + \nu_1 + \nu_2 + 1)}{\Gamma (j + \nu_2 + 1)
\Gamma (2k + \nu_1 + \nu_2 + 1) \Gamma (2k + \nu_1 + \nu_2 + 2)}
\nonumber\\
&\times&
\frac{\Gamma (J + k + \nu_1 + \nu_2 + 2) \Gamma (2j + \nu_2 + \nu_3 + 2)}{
\Gamma (J - j + \nu_1 + 1 ) \Gamma (j + k + \nu_2 + \nu_3 + 2)}
\frac{\cN (J, k | \nu_1, \nu_2, \nu_3)}{\cN
(J, j |\nu_3, \nu_2, \nu_1)} \nonumber\\
&\times&{_4F_3} \left( \left.
{
- J + k , J + k + \nu_1 + \nu_2 + \nu_3 + 2 , k + \nu_2 + 1 , k + 1
\atop
j + k + \nu_2 + \nu_3 + 2 , 2 k + \nu_1 + \nu_2 + 2, k - j + 1
}
\right| 1 \right) .
\end{eqnarray}
These coefficients are real $W_{jk} = W_{jk}^*$ and orthogonal
\begin{equation}
\sum_{\ell = 0}^{J} W_{j \ell} W_{k \ell} = \delta_{jk}.
\end{equation}
They satisfy the three-term recursion relation
\begin{eqnarray}
\label{RecRelW}
&&\left( j + \frac{\nu_2 + \nu_3}{2} \right)
\left( j + \frac{\nu_2 + \nu_3}{2} + 1 \right)
W_{jk} (J) \nonumber\\
&&\qquad\qquad\qquad=
{[ {\mbox{\boldmath$\hat J$}}^2_{23} ]}_{k,k}
W_{jk} (J)
+
{[ {\mbox{\boldmath$\hat J$}}^2_{23} ]}_{k,k - 1}
W_{j,k - 1} (J)
+
{[ {\mbox{\boldmath$\hat J$}}^2_{23} ]}_{k,k + 1}
W_{j,k + 1} (J)
\end{eqnarray}
with matrix elements of the Casimir operator
${\mbox{\boldmath$\hat J$}}^2_{23}$ evaluated above. Using it one can
easily derive useful approximate formula in the limits $\tau \equiv
\frac{k}{J} = {\rm fixed}$, $j = {\rm fixed}$, $J \to \infty$, to be
used later. At leading order in $J$ Eq.\ (\ref{RecRelW}) can be reduced
to the second order differential equation the solution to which is given
by
\begin{equation}
w_j (\tau) =
\sqrt{2}
\left[
\frac{\tau^{1 + 2 \nu_2} (1 - \tau^2)^{\nu_3} }{\omega_j (\nu_3, \nu_2)}
\right]^{1/2}
P_j^{(\nu_3, \nu_2)} (2 \tau^2 - 1),
\end{equation}
where\footnote{Here and below the powers of factor of $J$ in front of
functions in continuous limit are introduced in order to have
appropriate scaling properties for $J \to \infty$ when $\sum_{j = 0}^{J}
\to J \int_{0}^{1} d\tau$.} $W_{jk} = \frac{1}{\sqrt{J}} (- 1)^{J - k}
w_j (\tau)$.

\section{Three-site open spin chain and integrability.}

As we have established in the preceding sections the original
problem of calculation the spectrum of the evolution equation
(\ref{EvolEq}) is reduced in multicolour limit to the
diagonalization of the Hamiltonian (\ref{LargeNcHamiltonian})
which defines the one-dimensional open spin chain with three sites.
Moreover, the operators ${\mbox{\boldmath$\hat J$}}^2$ and
$\hat J^3$ are shown to be the integrals of motion of the system,
i.e.\ $[( {\mbox{\boldmath$\hat J$}}^2, \hat J^3 ), \cH]_- = 0$.
In addition to that there exists another `hidden' conserved charge
\cite{BraDerMan98} which thus leads to an exact integrability of
the three-particle system: the number of integrals of motion
equals the number of degrees of freedom. Let us show that the
system described by the Hamiltonian (\ref{LargeNcHamiltonian})
is equivalent to an integrable one-dimensional open spin chain
with impurities. For these purposes we will use the Quantum Inverse
Scattering Method (QISM) formalism. The latter has been used earlier
to prove the integrability of the multi-reggeon interaction in QCD
relevant for the high-energy behaviour of physical cross sections
and identify the corresponding model with the XXX Heisenberg magnet
of non-compact spin $s = 0$ \cite{Lip94,FadKor95}.

The central object in the algebraic operator formalism of the QISM
\cite{KorBogIze93,GomRuiSie96} is the $R$-matrix which depends
on a complex spectral parameter $\lambda$ and acts on the
tensor product of vector spaces $V_a \otimes V_b$ having the
dimensions of local spaces on each site. The indices $a$ and $b$
refer to the quantum and auxiliary spaces, respectively. The
$R$-matrix obeys the star-triangle Yang-Baxter equation
\cite{KorBogIze93,GomRuiSie96}
\begin{equation}
R_{a,b} ( \lambda - \mu ) R_{a,c} ( \lambda ) R_{b,c} ( \mu )
=
R_{b,c} ( \mu ) R_{a,c} ( \lambda ) R_{a,b} ( \lambda - \mu ).
\end{equation}

In our consequent considerations we only discuss the three-site open
spin chain. The monodromy matrix for it is defined as usual by
\begin{equation}
\label{Monodromy}
T_b ( \lambda ) = R_{a_1,b} ( \lambda - \delta_1 )
R_{a_2,b} ( \lambda - \delta_2 ) R_{a_3,b} ( \lambda - \delta_3 ),
\end{equation}
where the impurities, $\delta_i$, are introduced in order to enrich
the family of representations. The transfer matrix for open spin chain
\cite{Skl88}
\begin{equation}
\label{OpenTransfer}
\hat t_b ( \lambda )
= {\rm tr}_b\, K^+ ( \lambda ) T_b ( \lambda )
K^- ( \lambda ) T_b^{- 1} ( - \lambda ),
\end{equation}
with trace taken over the auxiliary space, --- has been shown by
Sklyanin to form a family of mutually commuting operators \cite{Skl88}
\begin{equation}
[ \hat t_{b_1} ( \lambda_1 ), \hat t_{b_2} ( \lambda_2 ) ]_- = 0 .
\end{equation}
This is a direct consequence of the Yang-Baxter equation. The boundary
matrices in Eq.\ (\ref{OpenTransfer}) satisfy the reflection Sklyanin
equations \cite{Skl88} and their simplest solution which leads to
the conformal invariant transfer matrix is given by the unit
matrix\footnote{Note that similar choice was made in the context
of open spin chains arisen in the problem of integrability of
quark-gluon reggeon equations \cite{KarKir99}.} $K^\pm = \1$.

For the two-dimensional auxiliary space (corresponding to spin
$\frac{1}{2}$) the $R$-matrix defines the Lax operator
\begin{equation}
L_a (\lambda) \equiv
R_{a,\frac{1}{2}} \left( \lambda - \frac{1}{2} \right)
= \lambda \1 + \sigma^i \hat J^i_a
=
\left(
\begin{array}{cc}
\lambda + \hat J^3_a & \hat J^-_a \\
- \hat J^+_a & \lambda - \hat J^3_a
\end{array}
\right) ,
\end{equation}
with $\hat J^\pm = \mp \left( \hat J^1 \pm i \hat J^2 \right)$.
The auxiliary transfer matrix which generates the set of local
integrals of motion can be redefined according to \cite{Skl88} as
\begin{equation}
\hat t_{\frac{1}{2}} \left( \lambda - \frac{1}{2} \right)
= \prod_{\ell = 1}^{3}
\Delta^{- 1} \left\{ R_{a_\ell,\frac{1}{2}}
( \lambda - \delta_\ell - 1 ) \right\}
t_{\frac{1}{2}} \left( \lambda \right),
\end{equation}
where $\Delta \left\{ R_{a_\ell,\frac{1}{2}} ( \lambda ) \right\}
= \lambda ( \lambda + 1 ) - {\mbox{\boldmath$\hat \jmath$}}^2_\ell$
stands for the quantum determinant \cite{KulSkl82,Skl92} and
\begin{equation}
t_{\frac{1}{2}} ( \lambda )
= {\rm tr}_{\frac{1}{2}}\,
L_{a_1} ( \lambda - \delta_1 ) L_{a_2} ( \lambda - \delta_2 )
L_{a_3} ( \lambda - \delta_3 )
\sigma_2
L_{a_3}^{\sf t} ( - \lambda - \delta_3 )
L_{a_2}^{\sf t} ( - \lambda - \delta_2 )
L_{a_1}^{\sf t} ( - \lambda - \delta_1 )
\sigma_2 ,
\end{equation}
which is an even function of the spectral parameter \cite{Skl88}
$t_{\frac{1}{2}} ( - \lambda ) = t_{\frac{1}{2}} ( \lambda )$.
The expansion of this expression w.r.t.\ the rapidity, $\lambda$,
defines the set of commuting integrals of motion.

For our purposes it is enough to restrict to the following values
of the shifts: $\delta_1 = \delta_3 \equiv \delta$, $\delta_2 = 0$.
An explicit calculation leads to the result
\begin{equation}
t_{\frac{1}{2}} ( \lambda )
= {\mit\Omega} ( \lambda )
- 4 \lambda^2
\left( \lambda^2 - {\mbox{\boldmath$\hat \jmath$}}^2_2 \right)
{\mbox{\boldmath$\hat J$}}^2
- 2 \lambda^2 \cQ ,
\end{equation}
where ${\mit\Omega} ( \lambda )$ is a c-number function
\begin{eqnarray*}
{\mit\Omega} ( \lambda )
= - 2
\left( \lambda^2 + {\mbox{\boldmath$\hat \jmath$}}^2_1 - \delta^2 \right)
\left( \lambda^2 + {\mbox{\boldmath$\hat \jmath$}}^2_2 \right)
\left( \lambda^2 + {\mbox{\boldmath$\hat \jmath$}}^2_3 - \delta^2 \right)
+ 4 \lambda^2 \left( \lambda^2 - \delta^2 \right)
\sum_{\ell = 1}^{3} {\mbox{\boldmath$\hat \jmath$}}^2_\ell
+ 4 \lambda^2
\left(
{\mbox{\boldmath$\hat \jmath$}}^2_1 {\mbox{\boldmath$\hat \jmath$}}^2_3
-
\delta^2 {\mbox{\boldmath$\hat \jmath$}}^2_2
\right),
\end{eqnarray*}
${\mbox{\boldmath$\hat J$}}^2$ is the total Casimir operator
(\ref{ThreeCasimir}) and $\cQ$ is a non-trivial integral of motion
\begin{equation}
\label{DeltaCharge}
\cQ
= [ {\mbox{\boldmath$\hat J$}}^2_{12},
{\mbox{\boldmath$\hat J$}}^2_{23} ]_+
- 2 \delta^2
\left\{
{\mbox{\boldmath$\hat J$}}^2_{12} + {\mbox{\boldmath$\hat J$}}^2_{23}
\right\}.
\end{equation}

The fundamental Hamiltonians which commutes with the above
set of charges can be constructed following Sklyanin \cite{Skl88}.
The fundamental monodromy matrix is defined by product
(\ref{Monodromy}) of $R$-matrices with equivalent representations
of conformal spin in quantum and auxiliary spaces and the shifts
of spectral parameters defined as above. Then the Hamiltonian
of the chain is defined as
\begin{equation}
\cH
= \frac{1}{2} \left. \frac{d}{d \lambda} \right|_{\lambda = 0}
\ln \hat t_b ( \lambda ) .
\end{equation}
Due to the diagonal form of the boundary matrices the latter is
given, up to irrelevant c-number term, by the sum of two-site
Hamiltonians \cite{Skl88}
\begin{equation}
\label{DeltaTotHamiltonian}
\cH = h_{a_1 a_2} + h_{a_2 a_3} ,
\end{equation}
which are expressed by the equation
\begin{equation}
h_{a b} =
\left. \frac{d}{d \lambda} \right|_{\lambda = 0}
\ln R_{a, b} ( \lambda - \delta )
=
R_{a, b}^{-1} ( - \delta ) R_{a, b}' ( - \delta ) .
\end{equation}
Making use of the explicit form of the Yang-Baxter bundle
\cite{KulResSkl81,Fad96}
\begin{equation}
R_{a,b} ( \lambda )
= P_{a b}\, \frac{\Gamma ( \hat J_{ab} + \lambda )}{
\Gamma ( \hat J_{ab} - \lambda )},
\end{equation}
one easily obtains
\begin{equation}
\label{DeltaHamiltonian}
h_{a b} = \psi ( \hat J_{a b} - \delta )
+ \psi ( \hat J_{a b} + \delta ) - 2 \psi ( 1 ) ,
\end{equation}
where we have legitimately added a c-number term to mimic the QCD
pair-wise Hamiltonian (\ref{LargeNcHamiltonian}).

The commutativity of $\cQ$ (\ref{DeltaCharge}) and $\cH$
(\ref{DeltaTotHamiltonian}) can be checked a posteriori by
an explicit calculation of the commutators $[\cQ, h_{(1,3)2}]_-$,
e.g.\ in the conformal basis (\ref{ThreePointBasis}). Here the
problem is reduced to the evaluation of the only matrix element
which reads
\begin{eqnarray*}
[\cQ, h_{12}]_{j, j + 1}
&=&
\left\{ [{\mbox{\boldmath$\hat J$}}^2_{12}]_{j + 1, j + 1}
+ [{\mbox{\boldmath$\hat J$}}^2_{12}]_{j,j}
- 2 \delta^2 \right\}
\left\{ [h_{12}]_{j + 1} - [h_{12}]_j \right\}
[ {\mbox{\boldmath$\hat J$}}^2_{23} ]_{j, j + 1} \\
&=& 2
\left\{ [{\mbox{\boldmath$\hat J$}}^2_{12}]_{j + 1, j + 1}
- [{\mbox{\boldmath$\hat J$}}^2_{12}]_{j, j} \right\}
[ {\mbox{\boldmath$\hat J$}}^2_{23} ]_{j, j + 1},
\end{eqnarray*}
where in the last line we have used the explicit form of the
matrix elements of the Hamiltonian (\ref{DeltaHamiltonian}).
This result together with similar calculation for $[\cQ, h_{23}]_-$
in $P_{13}$-transformed basis gives
\begin{equation}
[ \cQ, h_{12} ]_-
= 2 [ {\mbox{\boldmath$\hat J$}}^2_{23} ,
{\mbox{\boldmath$\hat J$}}^2_{12}]_- ,
\qquad
[ \cQ, h_{23} ]_-
= 2 [ {\mbox{\boldmath$\hat J$}}^2_{12} ,
{\mbox{\boldmath$\hat J$}}^2_{23} ]_- ,
\end{equation}
and thus $[ \cQ, \cH ]_- = 0$.

Finally, if we will put $\delta = \frac{3}{2}$ as suggested by QCD
(\ref{LargeNcHamiltonian}) we immediately obtain the conserved charge
`empirically' found in Ref.\ \cite{BraDerMan98}.

\section{Master equation.}

Thus as has been said above the existence of the additional charge
$\cQ$ leads to complete integrability of the quark-gluon system
($\nu_1 = \nu_3 = 1$, $\nu_2 = 2$). This implies that the
Hamiltonian $\cH$ is a function of the integrals of motion only,
${\mbox{\boldmath$\hat J$}}^2$ and $\cQ$, and allows to reduce the
complicated eigenvalue problem for the Hamiltonian to the more
simple one for the charge $\cQ_T$
\begin{equation}
\label{QTeigenEq}
\cQ_T {\mit\Psi} = q_T {\mit\Psi}
\qquad\mbox{with}\qquad
\cQ_T
= [ {\mbox{\boldmath$\hat J$}}^2_{12},
{\mbox{\boldmath$\hat J$}}^2_{23} ]_+
- \frac{9}{2}
\left\{
{\mbox{\boldmath$\hat J$}}^2_{12}
+ {\mbox{\boldmath$\hat J$}}^2_{23}
\right\} ,
\end{equation}
making use of the conformal invariance of the system. The latter
condition means that it is convenient to look for the solution
to (\ref{QTeigenEq}) in the form of expansion w.r.t.\ the three-point
basis (\ref{ThreePointBasis}) elaborated in the previous sections,
i.e.\
\begin{equation}
{\mit\Psi} = \sum_{j = 0}^{J} {\mit\Psi}_j \cP_{J;j} .
\end{equation}
The matrix elements of the charge possess, similar to the non-diagonal
Casimir operators ${\mbox{\boldmath$\hat J$}}^2_{(1,2)3}$, only three
non-zero diagonals
\begin{equation}
\label{QTmatrix}
\langle \cP_{J'; j'}
( \theta_1, \theta_2 | \theta_3 ) |
\cQ_T
| \cP_{J; j} ( \theta_1, \theta_2 | \theta_3 ) \rangle
= \delta_{J'J}
\left(
\delta_{j'j} {[ \cQ_T ]}_{j,j}
+
\delta_{j',j + 1} {[ \cQ_T ]}_{j + 1,j}
+
\delta_{j',j - 1} {[ \cQ_T ]}_{j - 1,j}
\right) ,
\end{equation}
where
\begin{eqnarray}
&&{[ \cQ_T ]}_{j,j}
=
2 \, [ {\mbox{\boldmath$\hat J$}}^2_{12} ]_{jj}
[ {\mbox{\boldmath$\hat J$}}^2_{23} ]_{jj}
- \frac{9}{2} \,
\left\{
[ {\mbox{\boldmath$\hat J$}}^2_{12} ]_{jj}
+ [ {\mbox{\boldmath$\hat J$}}^2_{23} ]_{jj}
\right\}, \\
&&{[ \cQ_T ]}_{j + 1,j}
=
\left\{
[ {\mbox{\boldmath$\hat J$}}^2_{12} ]_{j + 1,j + 1}
+ [ {\mbox{\boldmath$\hat J$}}^2_{12} ]_{jj}
- \frac{9}{2}
\right\} \,
[ {\mbox{\boldmath$\hat J$}}^2_{23} ]_{j + 1,j} ,
\end{eqnarray}
and ${[ \cQ_T ]}_{j + 1,j} = {[ \cQ_T ]}_{j,j + 1}$.
Therefore, the above equation (\ref{QTeigenEq}) leads to
a three-term recursion relation which can written in a concise
matrix form by introducing the two-dimensional vector
$\mbox{\boldmath${\mit\Psi}$}_j = \left( { {\mit\Psi}_j \atop
{\mit\Psi}_{j - 1} } \right)$ as
\begin{equation}
\label{RecRel}
\mbox{\boldmath${\mit\Psi}$}_{j + 1}
= \mbox{\boldmath$M$}_j
\mbox{\boldmath${\mit\Psi}$}_j
\qquad\mbox{with}\qquad
\mbox{\boldmath$M$}_j
=
\left(
\begin{array}{cc}
- \frac{{[ \cQ_T ]}_{j,j} - q_T}{{[ \cQ_T ]}_{j,j + 1}}
&
- \frac{{[ \cQ_T ]}_{j,j - 1}}{{[ \cQ_T ]}_{j,j + 1}}
\\
1 & 0
\end{array}
\right) .
\end{equation}
It must be supplied with the boundary conditions ${\mit\Psi}_{- 1}
= {\mit\Psi}_{J + 1} = 0$ which manifest the polynomial character
of the eigenfunctions. The solution to the Eq.\ (\ref{RecRel})
can be presented thus in the form
\begin{eqnarray*}
\mbox{\boldmath${\mit\Psi}$}_{j + 1}
= \mbox{\boldmath$M$}_j
\mbox{\boldmath$M$}_{j - 1} \dots
\mbox{\boldmath$M$}_0
\mbox{\boldmath${\mit\Psi}$}_0 ,
\end{eqnarray*}
with the `vacuum' state $\mbox{\boldmath${\mit\Psi}$}_0 =
\left( { 1 \atop 0 } \right)$ where we have voluntary assumed
${\mit\Psi}_0 = 1$ due to its arbitrary normalization. The boundary
condition ${\mit\Psi}_{J + 1} = 0$ presumes that the $q_T$ values are
the real roots of (yet unknown) orthogonal polynomials. For the
expansion coefficients themselves it implies that the eigenvalues of
the matrix $\mbox{\boldmath$M$}_j$ must be complex because otherwise
the solution will be a monotonic function of $j$ on the whole interval
$j \in [0,J]$ and the only way to fulfill the boundary condition is to
accept necessarily the trivial solution ${\mit\Psi}_j = 0$. Therefore,
from the eigenvalue equation, ${\rm det}[\mbox{\boldmath$M$} - M \cdot
\1] = 0$, we get the condition for existence of the non-trivial solution
to the recursion relation
\begin{equation}
\label{Existence}
\left( {[ \cQ_T ]}_{j,j} - q_T \right)^2
\leq 4 {[ \cQ_T ]}_{j,j + 1} {[ \cQ_T ]}_{j,j - 1}.
\end{equation}

Let us redefine the expansion coefficient according to
\begin{equation}
{\mit\Psi}_j \equiv \varrho_j {\mit\Upsilon}_j ,
\end{equation}
where
\begin{equation}
\varrho_j =
\left[
\frac{(j + 1)^3 (j + 3)^3}{(j + 2)^3} (J - j + 1) (J + j + 5)
\right]^{-\frac{1}{2}} .
\end{equation}
The recursion relation then takes more transparent form which reads
\begin{equation}
\label{MasterEq}
(2j + 3) {\mit\Upsilon}_{j + 1}
+
(2j + 5) {\mit\Upsilon}_{j - 1}
+
\varrho_j^2\, (2j + 3)(2j + 5)
\left( {[ \cQ_T ]}_{j,j} - q_T \right) {\mit\Upsilon}_j
= 0 .
\end{equation}
This master equation will be the main object of our consequent
analysis. Unfortunately, there is little hope to solve Eq.\
(\ref{MasterEq}) analytically for arbitrary values of the total
conformal spin, therefore, below we develop a WKB-type expansion
for the conserved charge and the energy of the three-site chain
at large $J$.

\section{Quantization of integral of motion.}

\begin{figure}[t]
\unitlength1mm
\begin{center}
\vspace{1.2cm}
\hspace{-2cm}
\begin{picture}(100,155)(0,0)
\put(0,60){\insertfig{12}{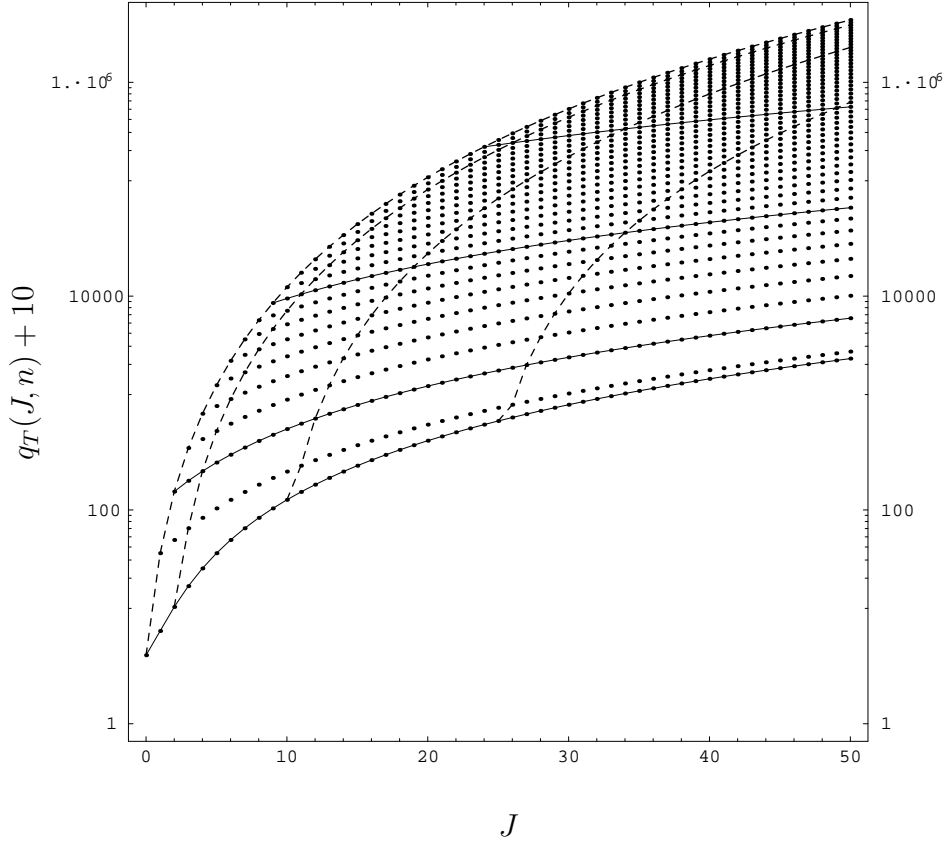}}
\put(60,60){$J$}
\put(-5,110){\rotate{$q_T (J, n) + 10$}}
\end{picture}
\end{center}
\vspace{-6.5cm}
\caption{\label{QTcharge} The spectrum of the conserved charge $q_T$
and a few trajectories from different sets (see the text).}
\end{figure}

From the constraint (\ref{Existence}) one can readily derive the critical
values of the spectrum of the conserved charge for asymptotical
values of the total conformal spin $J$. This leads to the inequality
\begin{equation}
\label{Limit}
0 \leq q_T / J^4 \leq \frac{1}{2} ,
\end{equation}
and to the fact that the maximum of the spectrum is achieved for
$j_{\rm max}= \frac{1}{\sqrt{2}} J$. This means that in the vicinity
of the upper boundary of the $q_T$-eigenvalues the coefficients
${\mit\Psi}_j$ are peaked around this point. The validity of the above
estimate can be checked by explicit numerical diagonalization of the
matrix (\ref{QTmatrix}) which gives the quantized values of the charge
$\cQ_T$. The resulting spectrum is shown in Fig.\ \ref{QTcharge}.
We can immediately notice that the values of $q_T$ form the families
of one-parametric curves. Obviously, there are two possible sets
of trajectories which apparently lead to an equivalent description
of the spectrum. The first set which starts from the top of the
spectrum (shown by dashed lines in Fig.\ \ref{QTcharge}) and behaves
as
\begin{eqnarray*}
q_T^{\rm top} = J^4
\left(
c^{\rm top}_0 (n) + c^{\rm top}_1 (n) J^{-1} + \cO (J^{-2})
\right).
\end{eqnarray*}
These trajectories are labeled by an integer $n$. Comparing this
expression with Eq.\ (\ref{Limit}) we conclude that $c^{\rm top}_0
(n) = \frac{1}{2}$. Another alternative is to adhere to the
parametrization of the curves starting from the bottom (shown by
solid curves)
\begin{eqnarray*}
q_T^{\rm bot} = J^2
\left( c^{\rm bot}_0 (m) + c^{\rm bot}_1 (m) [\ln J + a]^{- 2}
+ \cO (J^{-1}, \ln^{- 4} J) \right),
\end{eqnarray*}
labeled by $m$.

These results can be viewed as different analytical continuations of
the charge, defined for integer $J$, to complex values of the conformal
spin. Whatever choice we made, --- that can be judged by employing
external physical arguments, --- the result is the same: the fine
structure of the spectrum is generated by the pre-asymptotic corrections
in the conformal spin $J$.

\subsection{Description from below.}

Let us address first the question of the description of the spectrum
of the charge starting from below. In this case the trajectories behave
as $q_T \propto J^2$ and one can immediately find two exact solutions
to the master equation
\begin{equation}
\label{ExactQT}
q_T^{\rm exact-1} (J) = - \frac{53}{8} + (J + 1)^2,
\qquad
q_T^{\rm exact-2} (J) = - \frac{53}{8} + (J + 5)^2 ,
\end{equation}
separated by a gap from the rest of the spectrum. At this boundary
of $q_T$-values the classically allowed region, as deduced from
Eq.\ (\ref{MasterEq}), tends to spread over the whole interval
of attained values of $j$ with exception of small vicinities of the
end points $j_{\rm end} \sim 1,J$.

To find the WKB solution let us introduce the function ${\mit\Upsilon}_j
= J (-1)^j \phi (\tau)$ with $j \equiv \tau J$. Then the recursion
relation can be reduced in leading order in $J$ to the second order
differential equation for $\phi (\tau)$:
\begin{equation}
\tau^2 (1 - \tau^2) \phi'' (\tau) - \tau (1 - \tau^2) \phi' (\tau)
+ 2 (q_T^\star - 1 + 4 \tau^2) \phi (\tau) = 0,
\end{equation}
where $q_T^\star = J^{-2} q_T$. The solution to this equation
is given by
\begin{equation}
\label{WKBsolution}
\phi (\tau)
= C^{(+)} \tau^{1 + 2 i \eta_T} {_2 \cF_1} (i \eta_T | \tau^2 )
+ C^{(-)} \tau^{1 - 2 i \eta_T} {_2 \cF_1} (- i \eta_T | \tau^2 ) ,
\end{equation}
with $C^{(\pm)}$ being complex constants and
${_2 \cF_1} (\pm i \eta_T | \tau^2 )
=
{_2F_1} \left( \left.
{
\frac{3}{2} \pm i \eta_T ,\
- \frac{3}{2} \pm i \eta_T
\atop
1 \pm 2 i \eta_T
}
\right| \tau^2 \right)$, where we have introduced the shorthand
notation $\eta_T \equiv \frac{1}{2} \sqrt{2 q_T^\star - 3}$.

To determine the unknown coefficients $C^{(\pm)}$ one has to find
the solution of the master equation in the vicinity of the deflection
points $\tau \to 0,1$, i.e.\ $J \gg j \sim 1$ and $J \gg J - j
\sim 1$, respectively.

For $J \gg J - j \sim 1$ the master equation simplifies into
\begin{equation}
{\mit\Upsilon}_{j + 1}
+ {\mit\Upsilon}_{j - 1}
+ 2 {\mit\Upsilon}_{j} = 0,
\end{equation}
with boundary condition ${\mit\Upsilon}_{J + 1} = 0$, and the solution
to which is
\begin{equation}
{\mit\Upsilon}_{j} = (- 1)^j ( J - j + 1 ) .
\end{equation}
An important property of the solutions we have found above is that
there exists a region, $J - j \sim \sqrt{J}$, where both of them are
valid so that by sewing them together we get the unknown coefficients
$C^{(\pm)}$. Namely,
\begin{equation}
\label{CoefficientsC}
\frac{C^{(+)}}{C^{(-)}}
= - \frac{{_2 \cF_1}
(- i \eta_T | 1)}{{_2 \cF_1} (i \eta_T | 1)} .
\end{equation}

The quantization condition for the charge $q_T$ can be obtained
from the similar procedure used at the other end of values attained
by $j$. However, instead of an explicit solution of the resulting
reduced (but still quite complicated) recursion relation one can
adopt the strategy based on the definite symmetry properties of
the system under permutation of quark fields. Assuming the
eigenfunction of the `hidden' charge to be simultaneously the
eigenfunctions of the permutation operator $P_{13}$: $P_{13}
{\mit\Psi} = e^{i \varphi} {\mit\Psi}$ with $\varphi = 0, \pi$
and $P_{13} \cP_{J;j} (\theta_1, \theta_2| \theta_3) = \cP_{J;j}
(\theta_3, \theta_2| \theta_1)$; we get
\begin{equation}
{\mit\Psi}_j = e^{i \varphi}
\sum_{k = 0}^{J} W_{jk} (J) {\mit\Psi}_k .
\end{equation}
with the phase being the function of the conserved charge $q_T$ the
dependence on which is determined by the relation
\begin{equation}
\varphi = {\rm arg}\, (- 1)^J \sum_{j = 0}^{J}
(- 1)^j \frac{(j + 2)^3}{(j + 1)(j + 3)} {\mit\Upsilon}_j .
\end{equation}

Therefore, we can find the boundary solutions making use of known
WKB ones. Thus for the region $J \gg j \sim 1$ we get immediately
\begin{equation}
{\mit\Upsilon}_j = (- 1)^J
\left[ \frac{(j + 1)(j + 3)}{(j + 2)} \right]^{3/2} e^{i \varphi}
\int_{0}^{1}
d\tau \frac{w_j (\tau) \phi (\tau)}{[\tau^3 (1 - \tau^2)]^{1/2}} .
\end{equation}
After simple algebra it can be cast to the form
\begin{eqnarray}
\label{Boarder1}
{\mit\Upsilon}_j
\!\!\!&=&\!\!\!
(- 1)^{J + j} \frac{(j + 1)(j + 3)}{(j + 2)} e^{i \varphi}
\nonumber\\
&\times&\!\!\!
\left\{ C^{(+)} \left[ (- 1)^j {_2 \cF_1} (i \eta_T | 1)
+
\frac{\Gamma \left( \frac{3}{2} + i \eta_T \right)
\Gamma \left( \frac{5}{2} + j - i \eta_T \right)
}{\Gamma \left( \frac{3}{2} - i \eta_T \right)
\Gamma \left( \frac{5}{2} + j + i \eta_T \right)}
{_4 \cF_3} (i \eta_T | 1)
\right]
+ (+ \to -)
\right\}.
\end{eqnarray}
where ${_4 \cF_3} (i \eta_T | 1) =
{_4F_3} \left( \left.
{ - \frac{3}{2} + i \eta_T ,\ - \frac{1}{2} + i \eta_T ,\
\frac{3}{2} + i \eta_T ,\ \frac{3}{2} + i \eta_T
\atop
- \frac{3}{2} - j + i \eta_T ,\
\frac{5}{2} + j + i \eta_T ,\
1 + 2 i \eta_T } \right| 1 \right)$ and possesses
definite scaling properties when expressed by the series.
Moreover, Eq.\ (\ref{Boarder1}) obviously fulfills the boundary condition
${\mit\Upsilon}_{-1} = 0$.

From Eq.\ (\ref{WKBsolution}) it follows that for large $J$
\begin{equation}
{\mit\Upsilon}_j = (- 1)^j
\left\{
C^{(+)} j^{1 + 2 i \eta_T} J^{- 2 i \eta_T}
+
C^{(-)} j^{1 - 2 i \eta_T} J^{2 i \eta_T}
\right\}
\left\{ 1 + \cO (j/J) \right\} ,
\end{equation}
while from (\ref{Boarder1}) we have
\begin{eqnarray}
{\mit\Upsilon}_j
\!\!\!&=&\!\!\!
(- 1)^{J + j} e^{i \varphi}
\Biggl\{
C^{(+)} j^{1 - 2 i \eta_T}
\frac{\Gamma\left( \frac{3}{2} + i \eta_T \right)}{
\Gamma\left( \frac{3}{2} - i \eta_T \right)}
+
C^{(-)} j^{1 + 2 i \eta_T}
\frac{\Gamma\left( \frac{3}{2} - i \eta_T \right)}{
\Gamma\left( \frac{3}{2} + i \eta_T \right)} \nonumber\\
&&\qquad\qquad\quad +
(- 1)^j j
\left[
C^{(+)} {_2 \cF_1} (i \eta_T | 1)
+
C^{(-)} {_2 \cF_1} (- i \eta_T | 1)
\right]
\Biggr\}
\left\{ 1 + \cO (1/j) \right\} .
\end{eqnarray}
Matching these equations in the region of their overlap $j \sim \sqrt{J}$
making use of Eq.\ (\ref{CoefficientsC}) results into the following
quantization condition for the charge
\begin{equation}
2 \eta_T \ln J -
{\rm arg}\,
\frac{\Gamma \left( \frac{3}{2} + i \eta_T \right)}{
\Gamma \left( \frac{3}{2} - i \eta_T \right)}
\frac{{_2 \cF_1} (- i \eta_T | 1)}{{_2 \cF_1} (i \eta_T | 1)}
= \pi m,
\end{equation}
with $m \in \bfZ_+$. The trajectories generated by this equation
are the ones shown by solid lines in Fig.\ \ref{QTcharge}.

\subsection{Description from above.}

For the description from above the classically allowed region is
concentrated in the vicinity of the point $j_{\rm max} =
\frac{1}{\sqrt{2}} J$. Therefore, we expand the master equation
around it, $j = \frac{1}{\sqrt{2}} \left( J + \lambda \sqrt{J}\right)$,
and look for the solution to Eq.\ (\ref{MasterEq}) in the form of
series w.r.t.\ inverse powers of $J$
\begin{equation}
\label{WKBexp}
{\mit\Phi} (\lambda) = \sum_{\ell = 0}^{\infty}
{\mit\Phi}_{(\ell)} (\lambda) J^{- \ell/2} ,
\qquad\mbox{and}\qquad
q_T (J, n) = J^4 \sum_{\ell = 0}^{\infty} q_T^{(\ell)}(n) J^{- \ell},
\end{equation}
where ${\mit\Upsilon}_j \equiv {\mit\Phi} (\lambda)$.
This leads to the following sequence of differential equations
\begin{eqnarray}
\label{FirstEq}
&&\cD_{(1)} {\mit\Phi}_{(0)} (\lambda) = 0, \\
\label{SecondEq}
&&\cD_{(1)} {\mit\Phi}_{(1)} (\lambda)
+ \cD_{(2)} {\mit\Phi}_{(0)} (\lambda) = 0, \\
&&\cD_{(1)} {\mit\Phi}_{(2)} (\lambda)
+ \cD_{(2)} {\mit\Phi}_{(1)} (\lambda)
+ \cD_{(3)} {\mit\Phi}_{(0)} (\lambda) = 0, \nonumber\\
&&\cD_{(1)} {\mit\Phi}_{(3)} (\lambda)
+ \cD_{(2)} {\mit\Phi}_{(2)} (\lambda)
+ \cD_{(3)} {\mit\Phi}_{(1)} (\lambda)
+ \cD_{(4)} {\mit\Phi}_{(0)} (\lambda) = 0, \nonumber\\
&&\dots , \nonumber
\end{eqnarray}
supplied with boundary conditions ${\mit\Phi} (\pm\infty) = 0$
and explicit form of differential operators given by
\begin{eqnarray}
\cD_{(1)} \!\!\!&=&\!\!\! \frac{d^2}{d \lambda^2}
+ 4 \left( 6 - q^{(1)}_T - 2 \lambda^2 \right) ,
\nonumber\\
\cD_{(2)} \!\!\!&=&\!\!\! - \frac{d}{d \lambda}
+ 8 \lambda \left( 6 - 4 \sqrt{2} - \lambda^2 \right) ,
\nonumber\\
\cD_{(3)} \!\!\!&=&\!\!\! \frac{1}{6} \frac{d^4}{d \lambda^4}
+ 4 \left( 3 - \lambda^2 \right) \frac{d^2}{d \lambda^2}
+ \lambda \frac{d}{d \lambda}
- \left( 43 - 96 \sqrt{2} + 4 q^{(2)}_T - 24 \lambda^2
+ 48 \sqrt{2} \lambda^2 + 2 \lambda^4 \right) ,
\nonumber\\
\cD_{(4)} \!\!\!&=&\!\!\! - \frac{1}{3} \frac{d^3}{d \lambda^3}
+ 4 \lambda \left( 6 - 4 \sqrt{2} - \lambda^2 \right)
\frac{d^2}{d \lambda^2}
+ \left( - 12 + 2 \sqrt{2} + 3 \lambda^2 \right)
\frac{d}{d \lambda}
- 2 \lambda \left( 57 - 48 \sqrt{2} + 8 \sqrt{2} \lambda^2 \right) ,
\nonumber\\
\dots&&
\nonumber
\end{eqnarray}

The solution to Eq.\ (\ref{FirstEq}) is expressed by Hermite polynomials
\begin{equation}
\label{Hermite0}
{\mit\Phi}_{(0)} (\lambda)
= H_n \left( \sqrt{2 \sqrt{2}} \lambda \right) e^{- \sqrt{2} \lambda^2} .
\end{equation}
The solutions to others (\ref{SecondEq}) are given by
\begin{equation}
{\mit\Phi}_{(\ell)} (\lambda)
= \int d \lambda' G (\lambda, \lambda') \cJ_{(\ell)} (\lambda')
+ C_{(\ell)} {\mit\Phi}_{(0)} (\lambda),
\end{equation}
with the Green function of the homogeneous equation (\ref{FirstEq}),
$\cD_{(1)} G (\lambda, \lambda') = \delta (\lambda - \lambda')$,
and the source $\cJ_{(\ell)} = - \sum_{k = 0}^{\ell - 1}
\cD_{(\ell + 1 - k)} {\mit\Phi}_{(k)}$. Explicitly they
are given by a linear combination of the Hermite polynomials,
e.g.\ for the first non-trivial solution we have
\begin{equation}
\label{Hermite1}
{\mit\Phi}_{(1)} (\lambda)
= \frac{e^{- \sqrt{2} \lambda^2}}{2\sqrt{2\sqrt{2}}}
\sum_k c_k H_k \left( \sqrt{2\sqrt{2}} \lambda \right),
\end{equation}
with non-zero coefficients $c_{n + 3} = - \frac{1}{24}$, $c_{n + 1}
= - \frac{1}{4}( 33 - 24 \sqrt{2} + 3 n )$, $c_{n - 1} = n ( 17
- 12 \sqrt{2} + \frac{3}{2} n )$ and $c_{n - 3} = \frac{1}{3} n (n - 1)
(n - 2)$. Expressions for higher corrections are more involved.

The quantization for the charge $q_T^{(1)}$ immediately follows from
boundary conditions. The quantization of higher WKB corrections,
$q_T^{(\ell)}$ ($\ell \geq 2$), appears from the condition of absence
of zero-mode in the source terms. In this way we derive a few
next-to-leading corrections to the charge
\begin{eqnarray}
q_T^{(0)} (n)
&=& \frac{1}{2} , \nonumber\\
q_T^{(1)} (n)
&=& 6 - \frac{1}{\sqrt{2}} - \sqrt{2} n , \nonumber\\
q_T^{(2)} (n)
&=& \frac{1}{16} \left( 371 - 72 \sqrt{2} \right)
+ \frac{1}{8} \left( 11 - 72 \sqrt{2} \right) n
+ \frac{11}{8} n^2 ,\\
\dots\quad\ \ && \nonumber
\end{eqnarray}
(with $n =0,1,\dots$) which allow to describe the upper part of the
spectrum with a high accuracy. The integer $n$ has an obvious
interpretation as the number of nodes of the solutions in the
classically allowed region.

\section{Energy.}

The main advantage of having the integrability of the system is that
the eigenfunctions of integrals of motion simultaneously are the ones
for the Hamiltonian. By taking the average of the Hamiltonian w.r.t.\
its eigenfunctions ${\mit\Psi}$ we readily obtain
\begin{equation}
\label{EnergyAnalyticSquare}
\frac{2}{N_c} \cE (J, n) =
2 \left( \sum_{j = 0}^{J} \varrho_j^2\, {\mit\Upsilon}_j^2 \right)^{-1}
\sum_{j = 0}^{J} \epsilon(j)\,
\varrho_j^2\, {\mit\Upsilon}_j^2 - \frac{3}{2} ,
\end{equation}
with $\epsilon (j) = \psi (j + 1) + \psi (j + 4) + 2 \gamma_E$.

At the same time using the representation of the fundamental basis
in terms of the hypergeometric function (\ref{HyperBasis}) and taking
into account definite symmetry properties of the system under
permutation of quarks one can derive the equation linear in the
expansion coefficients
\begin{equation}
\label{EnergyAnalyticLinear}
\frac{2}{N_c} \cE (J, n) =
\left(
\sum_{j = 0}^{J} (- 1)^j \frac{(j + 2)^3}{(j + 1)(j + 3)}
{\mit\Upsilon}_j
\right)^{-1}
\sum_{j = 0}^{J} (- 1)^j \epsilon (j)
\frac{(j + 2)^3}{(j + 1)(j + 3)} {\mit\Upsilon}_j
+ \frac{1}{3}.
\end{equation}
Both of these formulae display the complementary aspects of the
formalism. When supplied by the solution of the master equation they
provide the exact solution of the three-body problem under study.

Thus in order to find the explicit form of the energy as a function
of the conserved charges one has to substitute the WKB solutions
found before into Eq.\ (\ref{EnergyAnalyticSquare}) or
(\ref{EnergyAnalyticLinear}).

\subsection{Description from below.}
\label{EnergyBelow}

The lowest trajectories corresponding to the solutions (\ref{ExactQT})
were known before \cite{BBKT97,KoiNis97,BelMul97}
\begin{eqnarray}
\label{ExactE}
\cE^{\rm exact-1} (J) = \frac{N_c}{2}
\left\{
2 \psi (J + 3) + 2 \gamma_E - \frac{1}{2} - \frac{1}{J + 3}
\right\}, \nonumber\\
\cE^{\rm exact-2} (J) = \frac{N_c}{2}
\left\{
2 \psi (J + 3) + 2 \gamma_E - \frac{1}{2} + \frac{3}{J + 3}
\right\} .
\end{eqnarray}
The first (second) energy corresponds to the anomalous dimension
of the genuine twist-3 part of the structure function $e$
($h_L$) in multicolour QCD since as has been found in the
references alluded to above the eigenfunctions corresponding
to Eq.\ (\ref{ExactE}) coincide with the coefficient function which
enter into the relation between $e$ and $h_L$ and corresponding
three-particle correlators (\ref{etoZ},\ref{hLtoZ}). Therefore,
the description from below seems to have more physical grounds to
be trusted as a true analytical continuation of the anomalous
dimensions to complex $J$ since otherwise it looks dubious that
the trajectories (\ref{ExactE}) pick out just one point for a given
$J$ from another set of `genuine' analytical functions discussed
below in the next section \ref{EnergyAbove}.

The remainder of the spectrum can be found by inserting the expression
for the eigenfunctions (\ref{WKBsolution}) with (\ref{CoefficientsC})
into Eq.\ (\ref{EnergyAnalyticLinear}). We obtain the following result
for the energy levels after the gap (cf. \cite{BraDerMan98})
\begin{equation}
\label{TopAfterGap}
\frac{2}{N_c} \cE (J, \eta_T)
= 2 \ln J + 4 \gamma_E
+ 2 \, {\rm Re} \, \psi \left( \frac{3}{2} + i \eta_T \right)
- \frac{3}{2}.
\end{equation}
As will be shown by numerical analysis this result is valid with
high accuracy for the whole spectrum.

\subsection{Description from above.}
\label{EnergyAbove}

To evaluate the energy for the top of the spectrum it is more
convenient to use the definition (\ref{EnergyAnalyticSquare}).
The energy can be evaluated consistently in WKB approximation as
expansion in inverse powers of the conformal spin $J$ as
\begin{equation}
\label{EnergyWKB}
{\cal E} (J, n) = \frac{N_c}{2}
\left\{
{\cal E}^{(0)} (J)
+ \sum_{\ell = 1}^{\infty} {\cal E}^{(\ell)}(n) J^{- \ell}
\right\} .
\end{equation}
Substituting the solutions (\ref{Hermite0}), (\ref{Hermite1}) into
(\ref{EnergyAnalyticSquare}) and equating the coefficients in front
of $J^{- \ell}$ we get the leading and non-leading corrections to the
energy
\begin{eqnarray}
\cE^{(0)} (J)
&=& 4 \ln J + 4 \gamma_E - 2 \ln 2 - \frac{3}{2} ,\\
\cE^{(1)} (n)
&=& 12 - \sqrt{2} - 2 \sqrt{2} n ,\\
\dots\quad\ \ && \nonumber
\end{eqnarray}
Thus at large $J$ the asymptotic behaviour of the energy of the
three-site open spin chain is given by
\begin{equation}
\frac{2}{N_c} \cE (J, q_T)
= \ln \left( q_T/2 \right) + 4 \gamma_E
- \frac{3}{2} + \cO (J^{-2}) .
\end{equation}
And again as before in Eq.\ (\ref{TopAfterGap}) the energy is a
function of the conserved charges of the problem.

\subsection{Comparison with numerical analysis.}

\begin{figure}[p]
\unitlength1mm
\begin{center}
\vspace{-2cm}
\hspace{0cm}
\begin{picture}(100,155)(0,0)
\put(0,60){\insertfig{9}{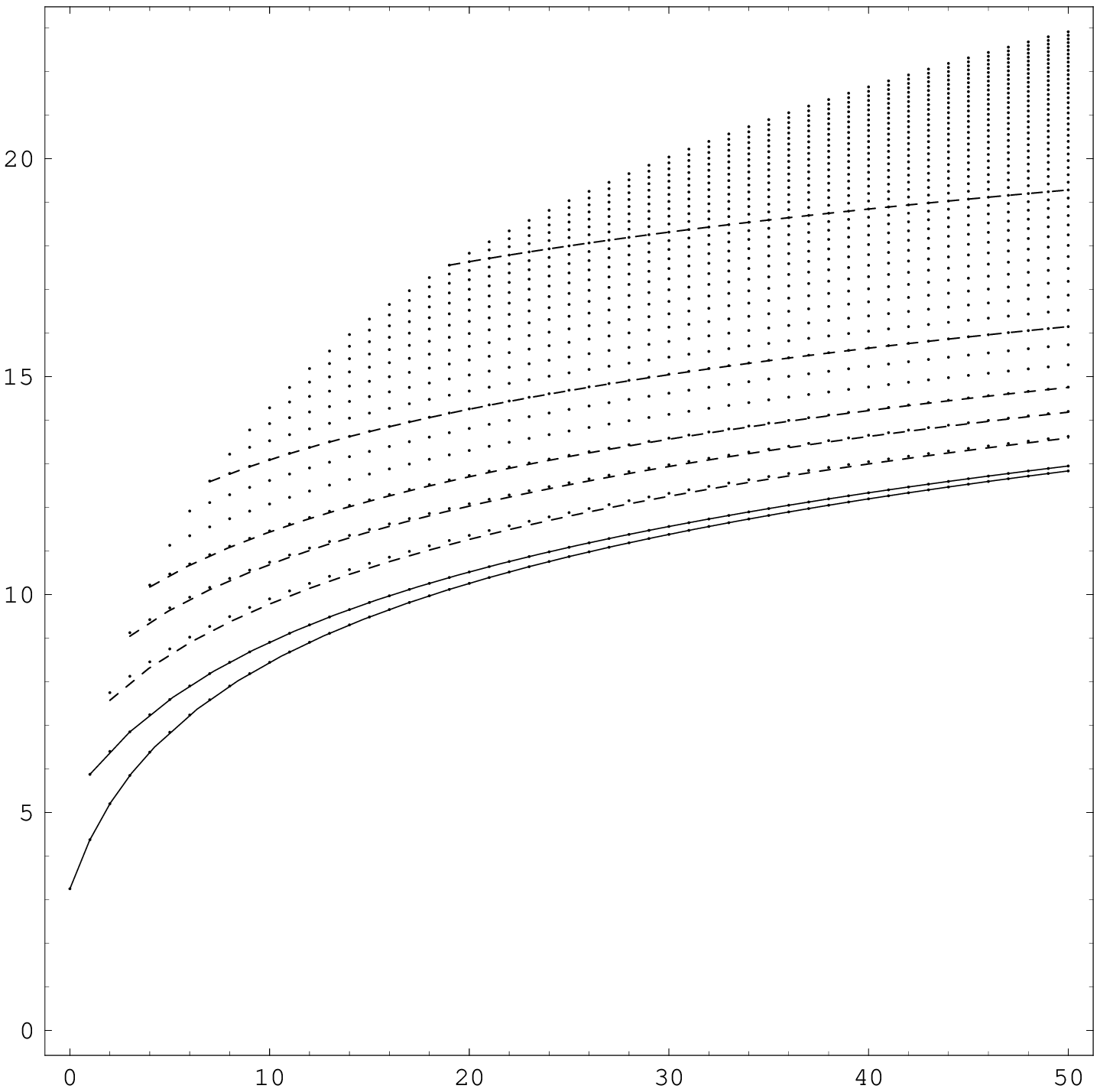}}
\put(46,55){$J$}
\put(-10,100){\rotate{${\cal E} (J, n)$}}
\end{picture}
\end{center}
\vspace{-6cm}
\caption{\label{EnergyBottom} The energy spectrum and the trajectories
(in dashed) defined by Eq.\ (\protect\ref{TopAfterGap}). The lowest two
eigenvalues (solid curves) are given by (\ref{ExactE}).}
\end{figure}

\begin{figure}[p]
\unitlength1mm
\begin{center}
\vspace{-2cm}
\hspace{0cm}
\begin{picture}(100,155)(0,0)
\put(0,60){\insertfig{9}{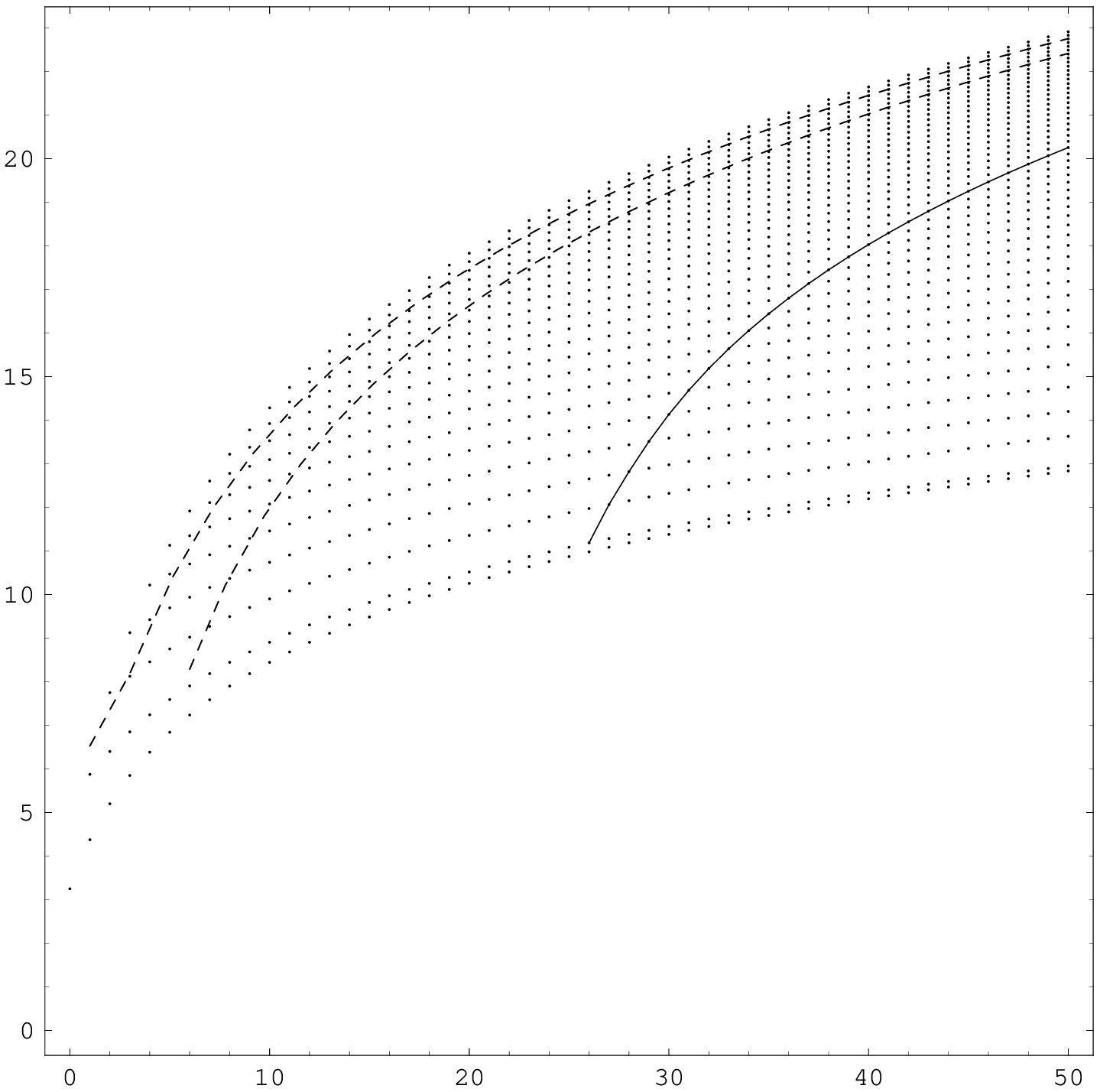}}
\put(46,55){$J$}
\put(-10,100){\rotate{${\cal E} (J, n)$}}
\end{picture}
\end{center}
\vspace{-6cm}
\caption{\label{EnergyTop} The spectrum of energy and a few
(dashed) trajectories described by the analytical formulae of
section \ref{EnergyAbove}.}
\end{figure}

Let us confront the analytical results derived so far with explicit
numerical diagonalization of the QCD anomalous dimension matrix.
This can be done either by using the expression of the Hamiltonian
in the conformal basis elaborated before
\begin{equation}
\frac{2}{N_c} \cH_{jk}
= \left( \epsilon (j) - \frac{3}{2} \right) \delta_{jk}
+ \sum_{\ell = 0}^{J} W_{\ell j} (J) \epsilon (\ell) W_{\ell k} (J),
\end{equation}
or making use of known expression for the multicolour anomalous dimension
matrix in the basis of ordinary local operators $\cZ_{J;j} \propto
\bar\psi (i \partial_+)^j \g G_{+\perp} (i \partial_+)^{J - j} \psi$
which is more suitable for numerical handling \cite{KoiNis97,BelMul97}
\begin{equation}
\frac{2}{N_c} \cH_{jk}
= \delta_{jk}
\left( \epsilon (j) - \frac{3}{4} \right) - \theta_{j, k + 1 }
\frac{( k + 2 )( k + 3 )}{( j - k )( j + 2 )( j + 3 )}
+ {j \rightarrow J - j \choose k \rightarrow J - k} .
\end{equation}
Here we have used the following step functions $\theta_{j,k}
= \{1,\ \mbox{if}\ j \geq k;\ 0,\ \mbox{if}\ j < k \}$.

The result of numerical diagonalization is presented in Figs.\
\ref{EnergyBottom} and \ref{EnergyTop} together with trajectories
derived from the study of the master equation. As we have mentioned
above the Eq.\ (\ref{TopAfterGap}) gives a perfect description for the
entire spectrum of the anomalous dimensions. While for the top of
the spectrum, due to limited number of terms taken in the WKB
expansion of the energy, we cannot penetrate too far inside the
spectrum since otherwise the condition $J \gg n$, --- whose validity
was assumed in the reduction of the recursion relation to the
set of differential equations (\ref{FirstEq},\ref{SecondEq}), ---
will be violated.

\section{Conclusions.}

Making use of the algebraic operator formalism of the Quantum Inverse
Scattering Method we have identified an integrable one-dimensional
inhohmogeneous three-site open quantum spin chain model to describe
the evolution of the twist-three chiral-odd quark-gluon correlators
in Quantum Chromodynamics in multicolour limit. The integrability
of the evolution equations results into substitution of the original
complicated problem of diagonalization of anomalous dimension matrix
to the more simple one for the `hidden' conserved charge. However,
since the problem turns out to be still quite complicated to derive
the analytical solution we have employed the WKB-type expansion w.r.t.\
the total conformal spin of the system. Taking the first few non-leading
corrections allows to describe the energy spectrum analytically with
good accuracy which can be systematically improved, however, at a price
of a rather involved analysis. The `hidden' charge unravel the structure
of the spectrum --- it distinguishes the particular components of
three-parton correlators with different renormalization group dependence.

Once the $1/N_c^2$ corrections are taken into account the integrability
of the system will be immediately violated. An interesting question is
how the trajectories of multicolour anomalous dimensions will mix
with each other once the $1/N_c^2$ effects are switched on.

The same techniques, up to minor modifications required due to the
loss of definite permutation symmetry of quark fields, can be applied
to the chiral-even sector corresponding the transverse spin structure
function $g_2 (x)$.

\vspace{1cm}

The author was supported by the Alexander von Humboldt Foundation.

\end{document}